\documentclass[twocolumn]{openjournal}

\shorttitle{Willamo et al.}
\shortauthors{Willamo et al.}

\begin{document}

\title{\object{V889 Her}: abrupt changes in the magnetic field or differential rotation?}

\email{email: teemu.willamo@helsinki.fi}

\author{Teemu Willamo}

\author{Thomas Hackman}
\affiliation{Department of Physics, P.O. Box 64, FI-00014 University of Helsinki, Finland}

\author{Jyri J. Lehtinen}
\affiliation{Finnish Centre for Astronomy with ESO (FINCA), University of Turku, Vesilinnantie 5, FI-20014 University of Turku, Finland}
\affiliation{Department of Physics, P.O. Box 64, FI-00014 University of Helsinki, Finland}

\author{Maarit J. Korpi-Lagg}
\affiliation{Department of Computer Science, Aalto University, PO Box 15400, FI-00076 Aalto, Finland}
\affiliation{Max Planck Institute for Solar System Research, Justus-von-Liebig-Weg 3, D-37077 G\"{o}ttingen, Germany}

\author{Oleg Kochukhov}
\affiliation{Department of Physics and Astronomy, Uppsala University, Box 516, 751 20 Uppsala, Sweden}

\begin{abstract}

  We have applied Zeeman-Doppler imaging (ZDI) to an extensive spectropolarimetric HARPSpol data set of the magnetically active young solar analogue V889 Her, covering 35 spectra obtained during six nights in May 2011. The data set allows us to study Stokes V profiles of the star at almost identical rotational phases, separated by one or more stellar rotations. We use these data to study if the line profiles evolve from one rotation to the next, and find that some evolution does indeed occur. We consider two possible explanations for this: abrupt changes in the large-scale magnetic field or differential rotation. We find it quite difficult to distinguish between the two alternatives using ZDI alone. 
A strong differential rotation could, however, explain the changes in the line profiles, so we conclude that it must be present, and the abrupt magnetic field evolution is left uncertain. Commonly, rapidly rotating stars are assumed to have only weak differential rotation. If the strong differential rotation of V889 Her is indeed present, as has been found in other studies as well, it could indicate that the theoretical and numerical results of differential rotation still need to be revised. The rapid changes that may have occurred in the magnetic field indicate that one should be quite cautious when interpreting ZDI maps constructed from data over long time intervals.

\end{abstract}

\keywords{stars: activity --- stars: individual (V889 Her) --- 
stars: magnetic field --- stars: solar-type}

\section{Introduction} \label{sec:intro}

\noindent Zeeman-Doppler imaging (ZDI) is a powerful tool used to indirectly map the magnetic field at the stellar surface \citep[e.g.][]{brown1991,kochukhov2016_zdi}. This is a further development of Doppler imaging \citep[DI;][]{vogt87,piskunov90}, which is used to map the surface temperature distribution. The magnetic field is mapped by inverting a series of Stokes I and V profiles (and in some cases also Q and U profiles) of absorption lines, distributed over the stellar rotation phases, thus finding the magnetic field and brightness distribution which reproduces the observed rotational changes in the line profiles. To map the whole visible stellar surface, a good coverage over the rotational phases is required. Depending on the stellar rotation period, this requires observations over many nights, especally for more slowly rotating stars or stars with a rotation period close to a small multiple of one day. In order to produce a sensible ZDI map, one has to assume that the magnetic field does not evolve during these observations. Similarly, with DI, it is assumed that the spot configuration stays constant through the observations. However, \cite{strassmeier2000_V711Tau} showed that there has been some rapid spot evolution happening in the RS CVn type star V711 Tau. \cite{strassmeier2003_V1192Ori} analyzed two DI maps from two consecutive rotations of the K giant V1192 Ori, and noted some differences, which they concluded to be due to anti-solar differential rotation. \cite{donati2016,donati2017} noted differences in the magnetic topology of two data sets of the T Tauri type star V830 Tau, which were separated by approximately one month, and were explained with differential rotation. Also \cite{lehtinen2022_LQHya} noted some evolution of the Stokes V profiles of the young solar analogue LQ Hya in two data sets with the duration of 9 and 11 consecutive nights.

The rate of changes happening in the magnetic field corresponds to the size of the magnetic structures. Large structures generally evolve slowly, while small structures can evolve more rapidly \citep{giles2017}. Small-scale fields, however, are difficult to detect with ZDI, since opposite polarities of nearby regions cancel each other out. At worst, ZDI is able to recover less than 1 $\%$ of the total magnetic field energy \citep{kochukhov2020}.

The object of this study, V889 Her, is a magnetically active \citep[$\log R'_{\mathrm{HK}} = -4.175$;][]{lehtinen16}, young solar analogue, of spectral class G2 V \citep{montes2001} and an estimated age of 30-50 million years \citep{strassmeier03}. There are a multitude of published DI and ZDI maps of V889 Her, most recently by \cite{willamo2019}. They published 19 temperature maps of V889 Her, where the main feature was a dark polar spot present in all of them. Previous DI maps have been published by \cite{strassmeier03}, \cite{jarvinen08}, \cite{huber09} and \cite{frasca10}, and ZDI maps by \cite{marsden06} and \cite{jeffers08}, who both found also a strong differential rotation to be present. \cite{huber09} found no evidence for differential rotation, and \cite{kovari11} found a differential rotation that was an order of magnitude weaker than that found by \cite{marsden06} and \cite{jeffers08}.

According to both theoretical \citep{kitchatinov1999} and numerical \citep[e.g.][]{viviani2018} studies, rapidly rotating stars should not have very strong differential rotation relative to their rotation period. Many observational studies also agree with this \citep[e.g.][]{reiners03,lehtinen16}, although, as seen with the conflicting results of V889 Her, the observational results are not as clear. In the study of \cite{reinhold2015} we can see that the differential rotation relative to the rotation rate decreases for rapid rotators, although the absolute rotational shear actually slightly increases. All in all, differential rotation has proven to be a difficult parameter to measure accurately from real stars \citep[see also e.g.][]{hackman2019}. Since differential rotation is a large unknown factor, it is unclear to what extent it can explain apparent evolution of spectral line profiles, and when these apparent changes correspond to true evolution of the magnetic field or spot structures.

\section{Data}

\noindent We analyze data of V889 Her obtained with the HARPS{-}pol high-resolution spectropolarimeter \citep{piskunov2011_harps} at the ESO 3.6m telescope at La Silla, Chile, retrieved from the ESO archives. The observations are from May 15--20, 2011. They were reduced with the REDUCE package \citep{piskunov02}. Spread over a period of 6 nights and containing a total of 35 individual spectra, this is a data set of remarkably high quality, in terms of the signal-to-noise ratio and phase coverage. See Table \ref{obs} for more details. The phase coverage was enhanced by combining only two sub-exposures, instead of the standard four, for the production of LSD profiles (see next Section). Due to this, no null profiles are available for the spectra.

To calculate the rotation phase $\phi$ of the star we have used the ephemeris:

\begin{equation}
\mathrm{HJD}_{\phi=0} = 2449950.550 + P_{\mathrm{rot}} \times E.
\end{equation}

\noindent The rotation period $P_{\mathrm{rot}} = 1.3300 \pm 0.0031$ d was derived using photometry from \cite{willamo2019}, including only the 28 data points between Julian Dates 2455685 and 2455710, corresponding to May 2011. This way we get the best period for this particular data set, and are not affected by fluctuations in the period due to differential rotation or other factors. We determined the best period with the Three Stage Period Analysis method \citep{jetsu1999_TSPA}.

\begin{table}
\centering
\caption{\label{obs}Observation log for V889 Her. \#Night numbers the observing nights as they are referred to in the text. $S/N$ is the signal-to-noise ratio of the Stokes LSD V-profile.}
\begin{tabular}{c c c c c}
\hline\hline
\#Night & Date (UT) & HJD-2455000 & $\phi$ & $S/N$\\
\hline
1 & 2011 05 15 5:49 & 696.743 & 0.446 & 41833\\
1 & 2011 05 15 6:17 & 696.763 & 0.461 & 40821\\
1 & 2011 05 15 7:51 & 696.828 & 0.510 & 43419\\
1 & 2011 05 15 8:19 & 696.847 & 0.524 & 44942\\
1 & 2011 05 15 9:42 & 696.905 & 0.568 & 37110\\
1 & 2011 05 15 10:04 & 696.920 & 0.579 & 36540\\
1 & 2011 05 15 10:27 & 696.936 & 0.591 & 36018\\
2 & 2011 05 16 5:50 & 697.744 & 0.198 & 42084\\
2 & 2011 05 16 6:18 & 697.763 & 0.213 & 42375\\
2 & 2011 05 16 7:50 & 697.827 & 0.261 & 42257\\
2 & 2011 05 16 8:18 & 697.847 & 0.276 & 44445\\
2 & 2011 05 16 9:47 & 697.908 & 0.322 & 42123\\
2 & 2011 05 16 10:15 & 697.928 & 0.337 & 42839\\
3 & 2011 05 17 5:19 & 698.722 & 0.934 & 41581\\
3 & 2011 05 17 5:47 & 698.742 & 0.949 & 44364\\
3 & 2011 05 17 8:18 & 698.847 & 0.028 & 39113\\
3 & 2011 05 17 8:40 & 698.862 & 0.039 & 37740\\
3 & 2011 05 17 10:05 & 698.921 & 0.083 & 28415\\
3 & 2011 05 17 10:26 & 698.936 & 0.095 & 27794\\
4 & 2011 05 18 5:23 & 699.725 & 0.688 & 33248\\
4 & 2011 05 18 5:51 & 699.745 & 0.703 & 37410\\
4 & 2011 05 18 8:17 & 699.846 & 0.779 & 37978\\
4 & 2011 05 18 8:38 & 699.861 & 0.790 & 38071\\
4 & 2011 05 18 9:00 & 699.876 & 0.802 & 36297\\
5 & 2011 05 19 4:54 & 700.705 & 0.425 & 34808\\
5 & 2011 05 19 5:21 & 700.724 & 0.439 & 31547\\
5 & 2011 05 19 7:55 & 700.831 & 0.520 & 35895\\
5 & 2011 05 19 8:23 & 700.850 & 0.534 & 32362\\
5 & 2011 05 19 8:52 & 700.870 & 0.549 & 34577\\
6 & 2011 05 20 5:19 & 701.722 & 0.189 & 24631\\
6 & 2011 05 20 5:47 & 701.742 & 0.205 & 26713\\
6 & 2011 05 20 7:17 & 701.804 & 0.251 & 23753\\
6 & 2011 05 20 7:45 & 701.824 & 0.266 & 18648\\
6 & 2011 05 20 9:31 & 701.897 & 0.321 & 21406\\
6 & 2011 05 20 9:59 & 701.917 & 0.336 & 25625\\
\hline
\end{tabular}
\end{table}

\section{Zeeman-Doppler imaging}

\noindent As the polarization signal in individual spectral lines, caused by the magnetic field, is too weak to be reliably detected, we use the Least-Square Deconvolution (LSD) method to enhance the signal. The method works by combining thousands of spectral lines into a mean LSD profile \citep{kochukhov2010_lsd}.
The line parameters are extracted from the VALD database\footnote{\url{http://vald.astro.uu.se/}} \citep{piskunov95,kupka99}. In the final profiles, 6545 lines between 3900-7100 $\mathrm{\AA}$ were used, with a mean wavelength 5111.589 $\mathrm{\AA}$  and a mean Land\'{e} $g$-factor 1.215.

We use the code inversLSD, developed by \cite{kochukhov2014_LSD}, to invert the ZDI maps. 
The stellar surface is divided into 1876 elements with approximately equal sizes, and the radial, meridional and azimuthal magnetic field components and the brightness is calculated for each.
Instead of modeling directly the magnetic field components, the magnetic field is represented with spherical harmonic expansions
, where higher order harmonics represent the magnetic field on smaller scales. 
For the number of spherical harmonic expansions,
\cite{hackman16} found that increasing maximum angular degree $\ell_{\mathrm{max}}$ to more than 5 did not significantly improve the fit to the Stokes V profiles. Also \cite{lehtinen2022_LQHya} noted, that although they used $\ell_{\mathrm{max}} = 20$, most of the magnetic energy was found in harmonics below $\ell \leq 6$. Both these studies analyzed stars with somewhat different values for $v \sin i$ than V889 Her, but still of the same order of magnitude, so these results should be fairly comparable to V889 Her. Similarly to \cite{willamo2022}, we have limited the solution to $\ell_{\mathrm{max}} = 10$ harmonics. 

The magnetic field and brightness inversions are performed simultaneously, so that brightness inhomogeneities are taken into account in the magnetic field inversion. This is necessary, since a lower brightness will lead to a weaker polarization signal, which would be difficult to tell apart from a weaker magnetic field, if the inversions were done independently of each other. For the magnetic field, high-order modes are suppressed by a regularization function, while for the brightness inversion, Tikhonov regularization is used to constrain the solution to distributions with a minimal gradient. Smoother variations are thus preferred over sharper ones. The regularization and other parameters used in the inversion, as well as the stellar parameters, are shown in Table \ref{param}. A more thorough description of the regularization can be found in \cite{rosen16}.

In our previous study of V889 Her \citep{willamo2019}, the inclination $i$ was set to 60$^\circ$, following \cite{marsden06}. However, from the light curve recently obtained by the TESS space telescope\footnote{Available at\\ \url{https://mast.stsci.edu/portal/Mashup/Clients/Mast/Portal.html}}, V889 Her seems to have at least one transiting exoplanet, which would indicate that the inclination is close to 90$^\circ$, if the rotation of the star is aligned with the planetary disk. Thus we set the inclination to a higher value, $i = 70^\circ$, estimated with the equation:

\begin{equation} \label{inc}
\sin i = \frac{P_{\mathrm{rot}}v\sin i}{2\pi R},
\end{equation}

\noindent where $P_{\mathrm{rot}}$ is the rotation period, $v\sin i$ the projected rotation velocity, and $R$ the stellar radius. References for these parameters are given in Table \ref{param}.

\begin{table*}
\centering
\caption{Stellar parameters of V889 Her and parameters used in the ZDI inversions.}
\label{param}
\begin{tabular}{c c c}
\hline\hline
Parameter & Value & Reference \\
\hline
$P_{\mathrm{rot}}$ & 1.3300 d & This study \\
$v\sin i$ & 38.5 km/s & \cite{willamo2019} \\
$i$ & $70^\circ$ & This study \\
$R$ & 1.09 $R_\odot$ & \cite{strassmeier03} \\
Microturbulence & 1.6 km/s & \cite{jarvinen08} \\
Macroturbulence & 3.0 km/s & \cite{strassmeier03}\\
Regularization (brightness) & $5 \times 10^{-9}$ & \ldots \\
Regularization (magnetic field) & $8 \times 10^{-13}$ & \ldots \\
Effective Land\'{e} $g$ factor & 1.215 & \ldots \\
Mean $\lambda$ & 5630.0 Å & \ldots \\
\hline
\end{tabular}
\end{table*}

\section{Differential rotation}

\noindent The latitudinal differential rotation parameter $\alpha$ is defined as:

\begin{equation}
\alpha = \frac{\Omega_{\mathrm{eq}} - \Omega_{\mathrm{pol}}}{\Omega_{\mathrm{eq}}},
\end{equation}

\noindent and the absolute latitudinal rotational shear is:

\begin{equation}
\Delta\Omega = \Omega_{\mathrm{eq}} - \Omega_{\mathrm{pol}},
\end{equation}

\noindent where $\Omega_{\mathrm{eq}}$ and $\Omega_{\mathrm{pol}}$ are the angular velocities at the equator and at the poles, respectively. Published values for the differential rotation of V889 Her are $\Delta\Omega=0.402\pm0.044$ rad/d \citep{marsden06}, $\Delta\Omega=0.52\pm0.04$ rad/d for Stokes I data and $\Delta\Omega=0.47\pm0.04$ rad/d for Stokes V data \citep{jeffers08}, and $\Delta\Omega=0.042$ rad/d \citep{kovari11}. \cite{jarvinen08} found a weaker differential rotation than \cite{marsden06}, although they did not derive a quantitative value for it, and \cite{huber09} found no evidence for differential rotation. In \cite{willamo2019}, V889 Her was assumed to have no differential rotation. 

The differential rotation law is assumed to follow the solar differential rotation law, with the higher order terms neglected. The dependence of the surface rotation rate on the latitude $\theta$ will thus be:

\begin{equation}
\Omega(\theta) = \Omega_{\mathrm{eq}}(1 - \alpha \sin^2(\theta)).
\end{equation}

\noindent The differential rotation does not need to necessarily follow the solar differential rotation law, but it is assumed, since we have no information to make reliable guesses about alternative differential rotation laws.

With different values of $\alpha$, the rotation period $P_{\mathrm{rot}}$ must also be changed. Since $P_{\mathrm{rot}}$ in the model is measured at the equator, but the observational value comes from the latitude of the dominating spot structure, the optimal value of $P_{\mathrm{rot}}$ is expected to decrease with increasing $\alpha$. In practice, the best values for $\alpha$ and $P_{\mathrm{rot}}$ are found by trying a grid of different values, and finding the minimum for the deviation between model and observations.

\section{Stokes V profiles}

\noindent The data set analyzed here contains 35 spectra. A typical (Z)DI data set contains perhaps a little more than 10 spectra \citep[e.g.][]{willamo2019,hackman2019,lehtinen2022_LQHya,willamo2022}, which, if spread evenly over the rotational phases, already provide a sufficient phase coverage. There are also studies with similar amounts of phases as here, though, for example of V830 Tau \citep{donati2017}. Our data set of V889 Her also contains some spectra at almost identical phases, which are separated by one or a few rotations. These are essential for the aim of this study, since, if no magnetic field evolution or differential rotation is present, the surface structure seen to the observer, and thus also the line profiles, from these phases should be identical. These multiple observations of the same phase can thus be used as a rough check of the assumption of the stability of the magnetic field over the time span of the observations. They are, however, not very useful for improving the phase coverage, and are thus commonly avoided in the observations due to time limitations, especially when one tries to get sufficient observations of as many stars as possible. 

\begin{figure*}
   \centering
   \includegraphics[bb=30 50 400 600,width=8cm]{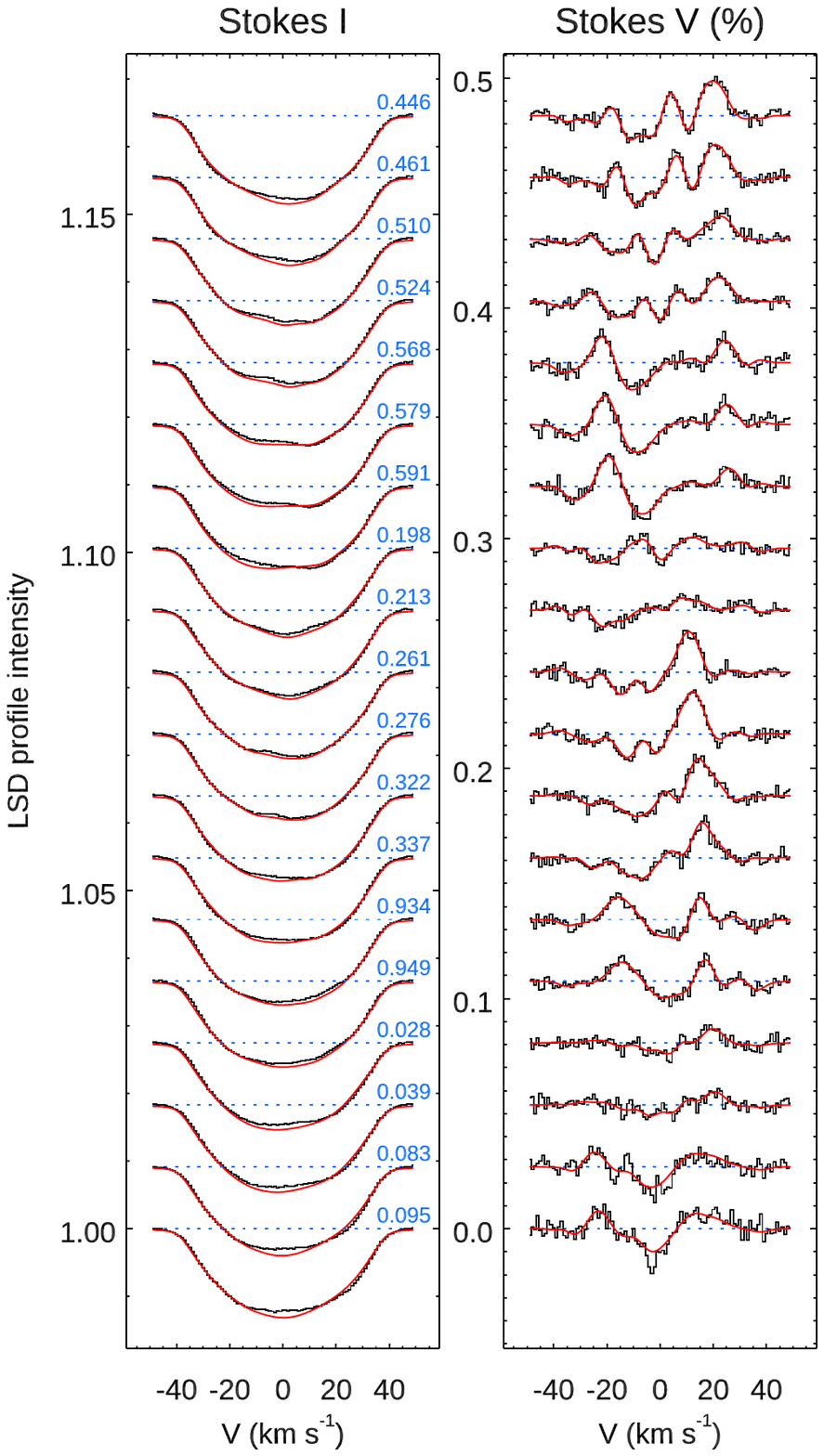}
   \includegraphics[bb=30 50 400 600,width=8cm]{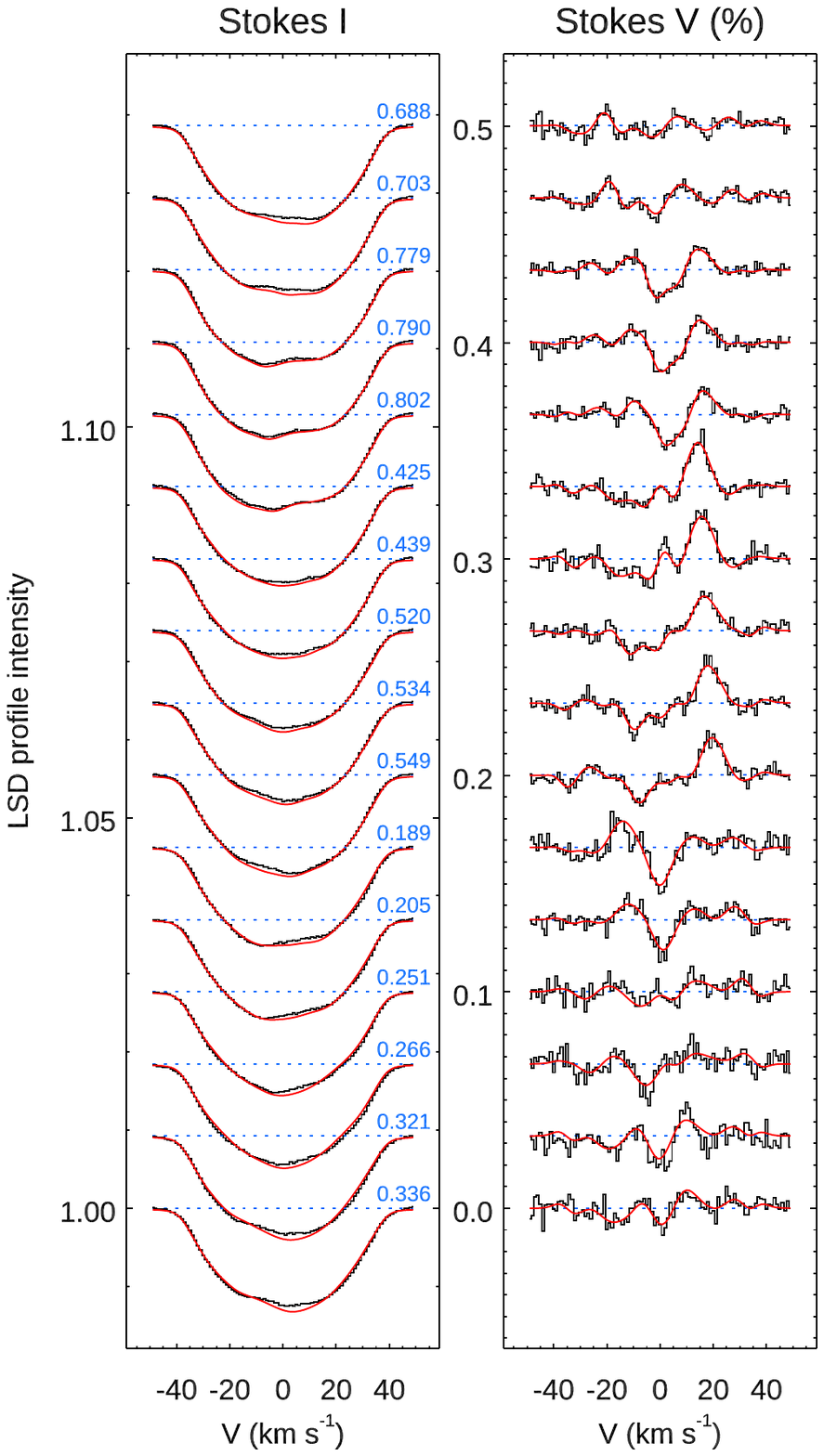}
   \includegraphics[width=10cm]{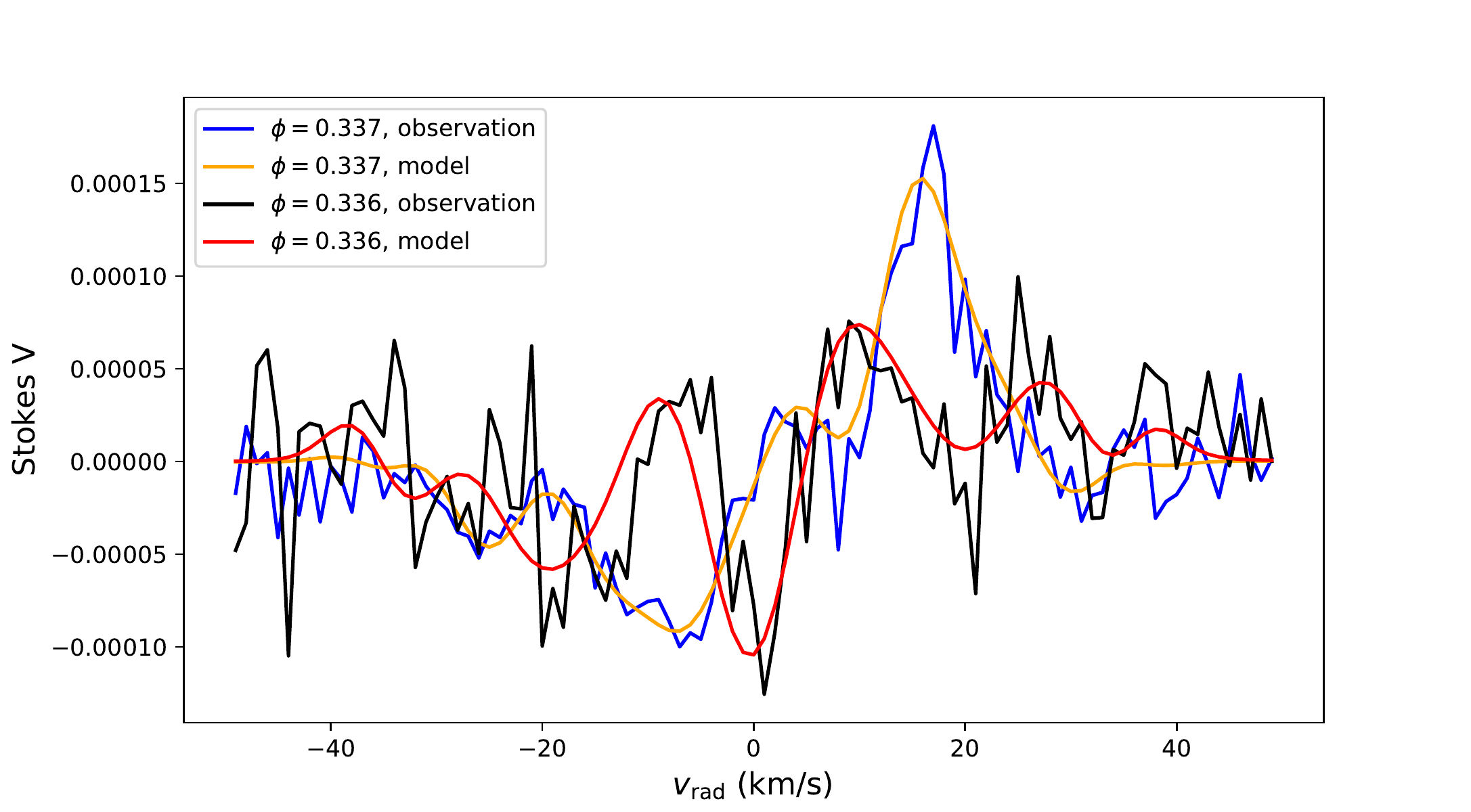}
   \caption{Upper panel: the Stokes LSD profiles for the three first nights (left) and the three last nights (right). The black lines are the observed LSD profiles, and the red lines are fits made to these. The rotational phase is indicated for each profile. Lower panel: V profiles for the two phases closest to each other, where the difference is most notable. The phase $\phi = 0.337$ is part of the map for nights 1-3, while the phase $\phi = 0.336$ is part of the map for nights 4-6, in Fig. \ref{maps}. This indicates that either a strong differential rotation must be present, or there has been some rapid evolution of the magnetic field.}
   \label{profile}
\end{figure*}

Fig.~\ref{profile} shows the Stokes LSD I and V profiles of the spectra, and the model fitted to the data, both separately for the first three nights and last three nights. There is especially one pair of profiles at $\phi = 0.337$ and $\phi = 0.336$, where the Stokes V profiles are clearly not identical, although separated only by $\Delta \phi = 0.001$ (shown together in the lower panel of Fig.~\ref{profile}). The difference between these is evidence that the star cannot have rotated as a solid body with a constant magnetic field structure during these observations, assuming that the rotation period we use is correct. Also phases around $\phi=0.2$ (0.198 and 0.213 in the first three nights, 0.189 and 0.205 in the last three nights) and $\phi=0.5$ (0.510 and 0.524 in the first three nights, 0.520 in the last three nights) are notably different in the figures.

Active stars, especially M dwarfs, are often showing flaring activity. Whereas flares can affect Stokes I profiles, they should not significantly affect the polarized Stokes V profiles \citep[e.g.][]{donati2017}. As a G-type star, V889 Her is not expected to show very high levels of flaring activity. 
A flare would also raise the continuum level of the I profile, which would be seen as a significant weakening of the line. This is not seen in any of the spectra. Thus, flares cannot explain the discrepancies seen at simultaneous rotational phases, as there are multiple examples of differences in the Stokes V profiles of closeby phases.

\section{Zeeman-Doppler maps}

\noindent In this section we present the ZDI maps constructed from our data. We first study the case where the star is assumed to rotate as a solid body with no differential rotation and the magnetic field is allowed to evolve between two subsets of the observations. After this we require the magnetic field to stay constant through the observations, and allow for differential rotation. These alternatives are then compared to each other, using the deviation between the model and observations: 

\begin{equation}
\sigma = \frac{\sum_{\phi,\lambda} \omega_{\phi,\lambda} (I_{\mathrm{obs},\phi,\lambda}-I_{\mathrm{mod},\phi,\lambda})^2}{n_\phi n_\lambda}, 
\end{equation}

\noindent where $I_{\mathrm{obs},\phi,\lambda}$ and $I_{\mathrm{mod},\phi,\lambda}$ are the observed and modeled intensities at each phase $\phi$ and wavelength $\lambda$, $\omega_{\phi,\lambda}$ is the weight of each data point, which is proportional to the square of the $S/N$ of the spectrum, and $n_\phi$ and $n_\lambda$ are the number of phases and wavelength points. $\sigma$ is shown for each model in Table \ref{devs}, where we see that the static model with no magnetic field evolution or differential rotation clearly has the highest deviation. 

\subsection{Case 1: Abruptly evolving magnetic field}

\noindent First we perform the analysis without differential rotation, but allowing the magnetic field to evolve during the observations.

As it was clear from Fig.~\ref{profile} that without differential rotation the magnetic field had to evolve during the observations, we divided our observation set in two, to get two subsets containing three consecutive nights each, labeled as \textquoteleft 1-3' 
and \textquoteleft 4-6\textquoteright. We also divided a third subset from the data, from which we removed the spectra of the first and last night, thus containing four nights from the middle of the observation set, labeled as \textquoteleft 2-5\textquoteright. It might be questionable whether the magnetic field stays constant even for the period of three or four nights, but by producing separate maps for these subsets, we show that the resulting fields are indeed different.

If we assume one spectrum to cover 10 \% of the rotation period, then all six nights combined have a phase coverage $f_{\phi} = 0.967$. With the division of the observations to three subsets, we have phase coverages $f_{\phi,1-3} = 0.744$ for the first set, $f_{\phi,2-5} = 0.925$ for the middle set and $f_{\phi,4-6} = 0.674$ for the last set, all being acceptable phase coverages on their own. We used the same inversion parameters for each subset.

\begin{figure*}
  \centering
   \includegraphics[bb=30 200 800 340,width=\textwidth,angle=180]{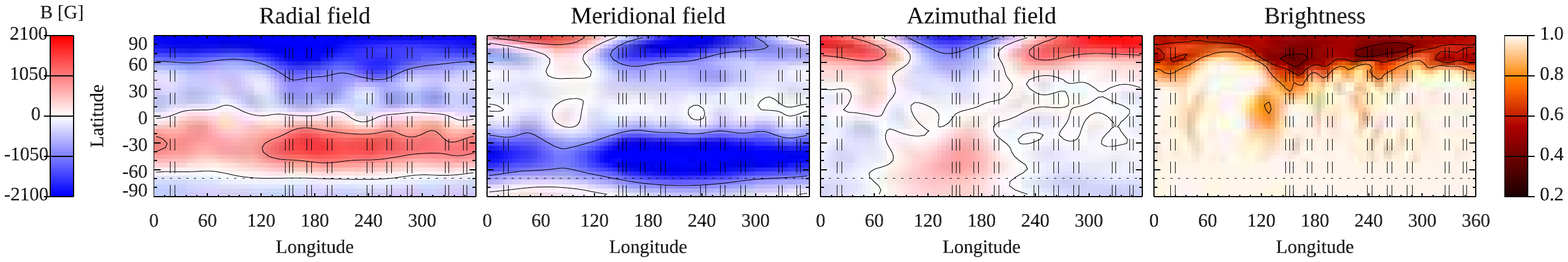}
   \includegraphics[bb=30 200 800 340,width=\textwidth,angle=180]{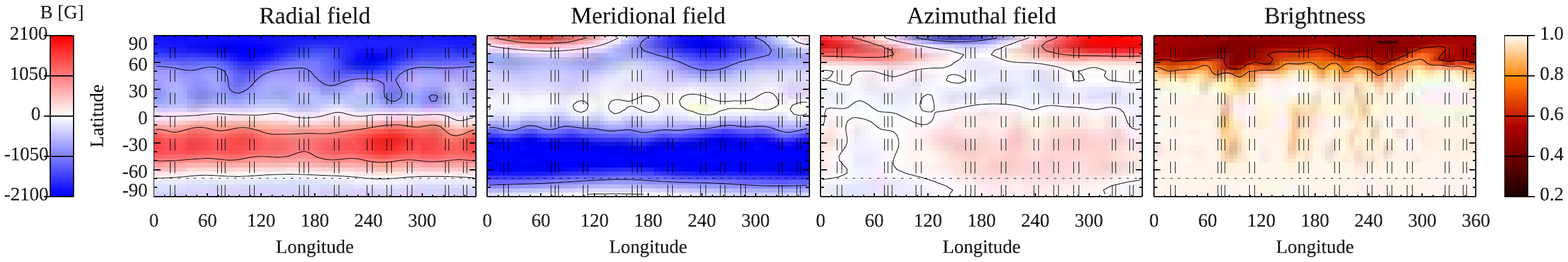} 
   \includegraphics[bb=30 200 800 340,width=\textwidth,angle=180]{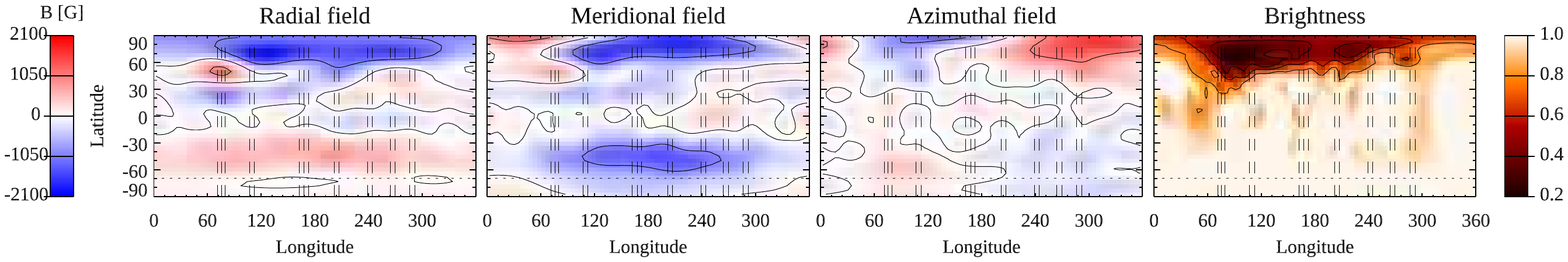}
   \caption{Magnetic field and brightness maps for nights 1-3 (up), 2-5 (centre) and 4-6 (down), without differential rotation. The vertical lines correspond to phases used for each map. The horizontal line indicates the limit of the visible surface due to the inclination.}
    \label{maps}
\end{figure*}

The three resulting ZDI maps are shown in Fig.~\ref{maps}, separately for the radial, meridional and azimuthal field components, and the brightness. 
We see that the subsets 1-3 and 2-5 are very similar, but differ from the subset 4-6, where the magnetic field strength is reduced significantly. If this is the case, then the magnetic field evolution would have happened quite abruptly during the last nights.

The deviation $\sigma$ for each of these maps is shown in Table \ref{devs}. To study whether an abruptly changing magnetic field or differential rotation is a better explanation to the data, these values should be compared to the one resulting from the analysis involving differential rotation, which is studied in the next section.

\subsection{Case 2: Differential rotation}

\begin{figure}
  \includegraphics[width=9cm]{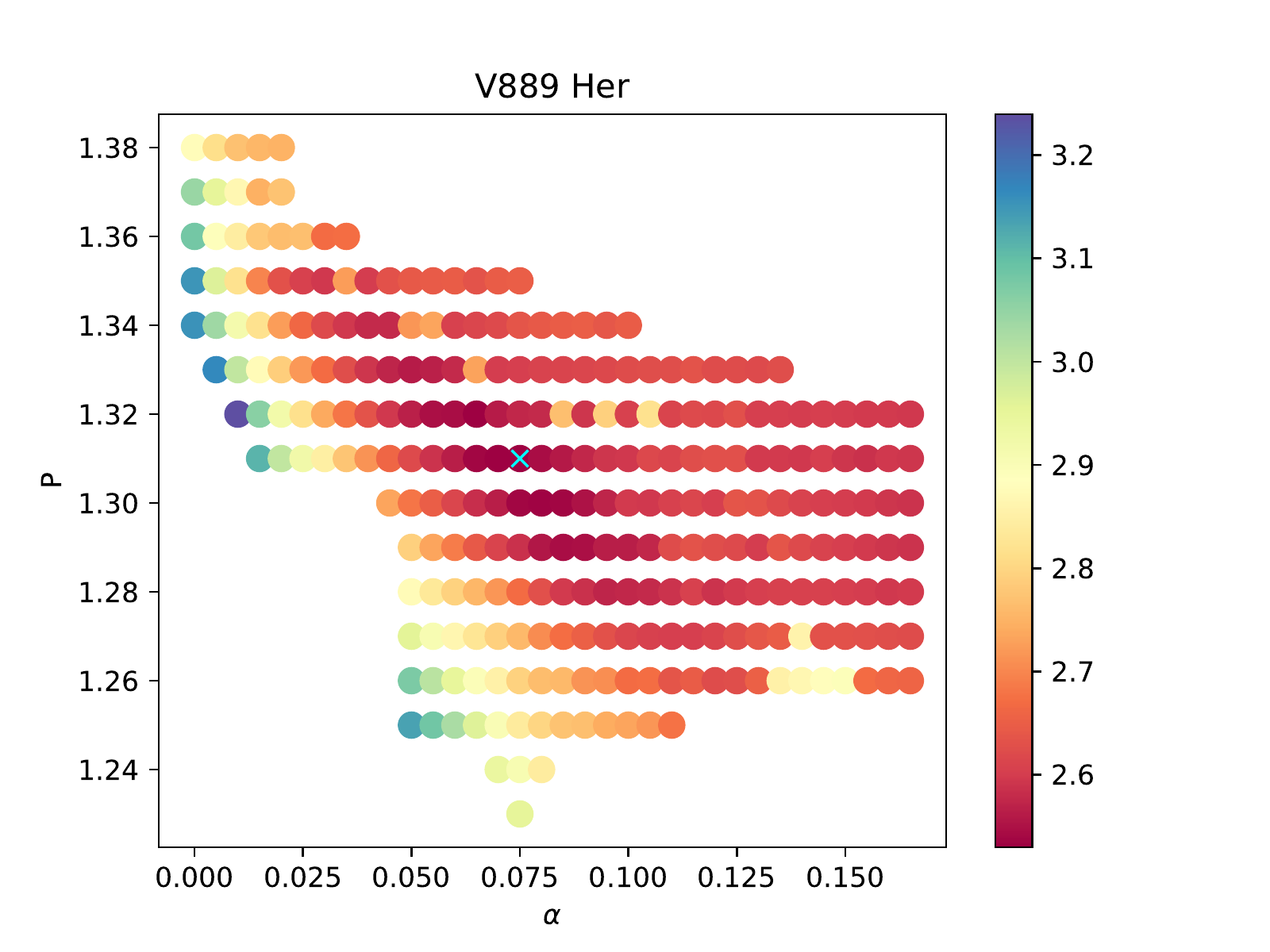}
  \caption{Map of the deviation $\sigma$ for differential rotation parameters of V889 Her. The colour scaling corresponds to the deviation between the model and observations in the respective ZDI map, with the unit being $10^{-5}$. The best fit is indicated with a cross.}
  \label{chi2}
\end{figure}

\begin{figure*}
  \centering
   \includegraphics[bb=30 200 800 340,width=\textwidth,angle=180]{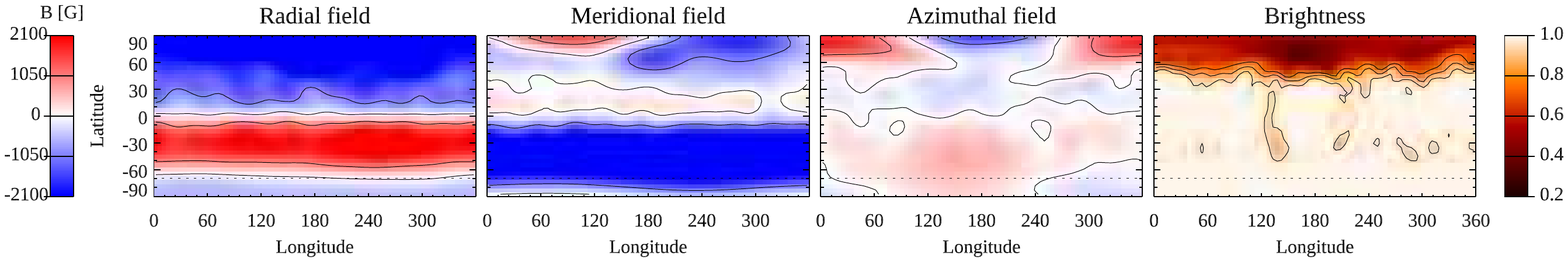}
   \caption{Magnetic field and brightness maps for all six nights with differential rotation parameters $\alpha = 0.075$ and $P_{\mathrm{rot}}=1.31$ d. The observed phases are not indicated, since with differential rotation there is no unique rotation period. The horizontal line indicates the limit of the visible surface due to the inclination. The maps are shown for the epoch HJD=2455699.0}
    \label{maps_difrot}
\end{figure*}

\begin{figure*}
    \centering
    \includegraphics[bb=30 50 400 600,width=8cm]{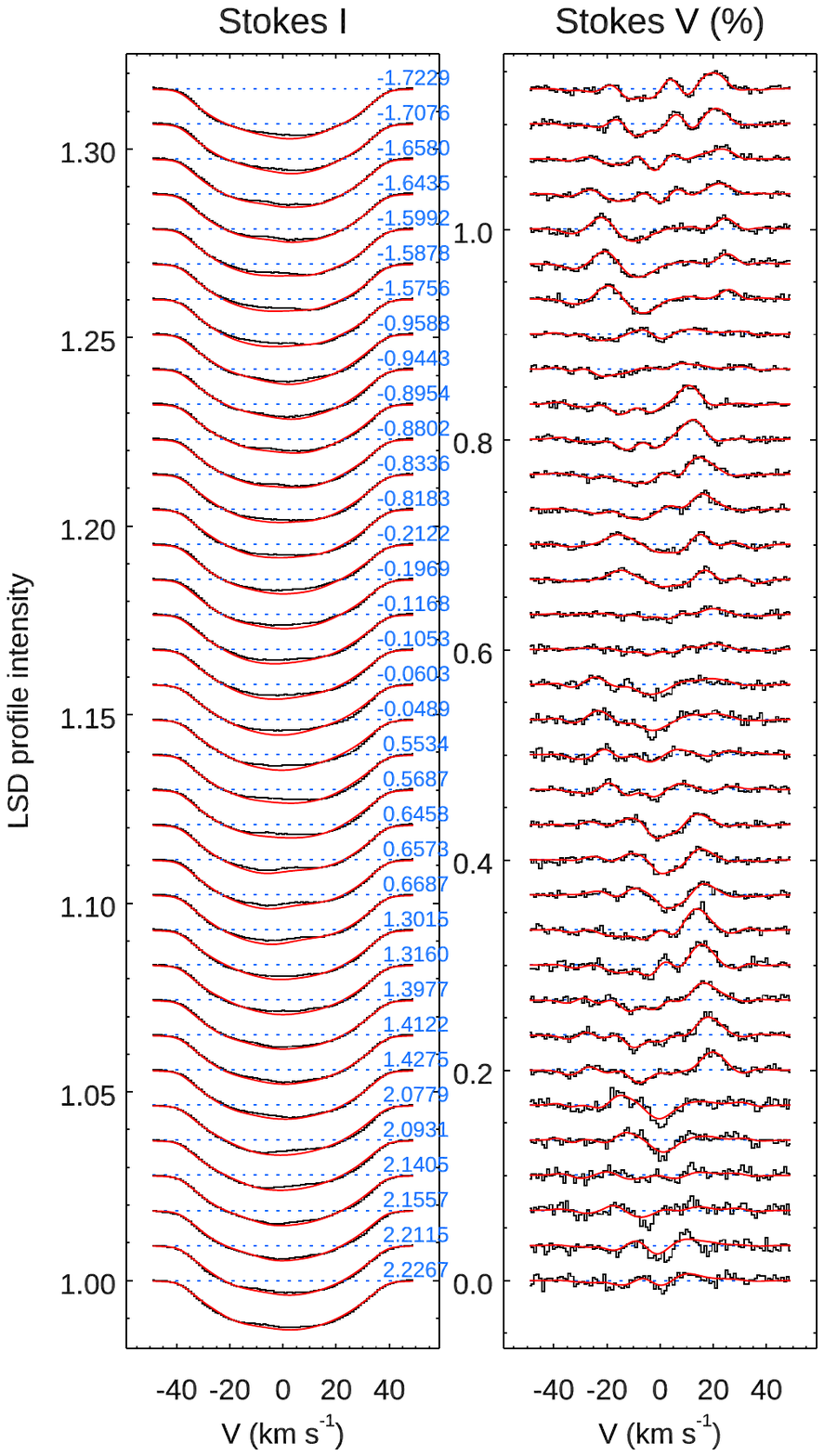}
    \caption{The Stokes profiles for the best value for differential rotation, $\alpha=0.075$ and $P_{\mathrm{rot}}=1.31$ d. The black lines are the observed LSD profiles, and the red lines are fits made to these. The rotational phase is indicated as HJD relative to 2455699.0}
    \label{prof_difrot}
\end{figure*}

\noindent Now we test how well the data can be reconstructed when differential rotation is allowed. This will stretch the magnetic field, which is assumed to be frozen into the stellar surface, but it is otherwise kept constant. A map of the deviation $\sigma$ for different values of $\alpha$ and $P_{\mathrm{rot}}$ is shown in Fig.~\ref{chi2}. The minimum for $\sigma$ is found with parameters $\alpha=0.075$\footnote{This corresponds to $\Delta\Omega = 0.36$ rad/d, slightly less than the values derived by \cite{marsden06} and \cite{jeffers08}.} and $P_{\mathrm{rot}} = 1.31$ d, with $\sigma = 2.5298 \times 10^{-5}$. The ZDI map corresponding to these best fit parameters is shown in Fig.~\ref{maps_difrot}. The corresponding LSD profiles are shown in Fig.~\ref{prof_difrot}. When looking at the profiles, the fits seem fairly good, which could mean that a strong differential rotation can explain the observed apparent line profile evolution even without significant abrupt changes in the magnetic field.

\begin{figure}
    \centering
    \includegraphics[width=9cm]{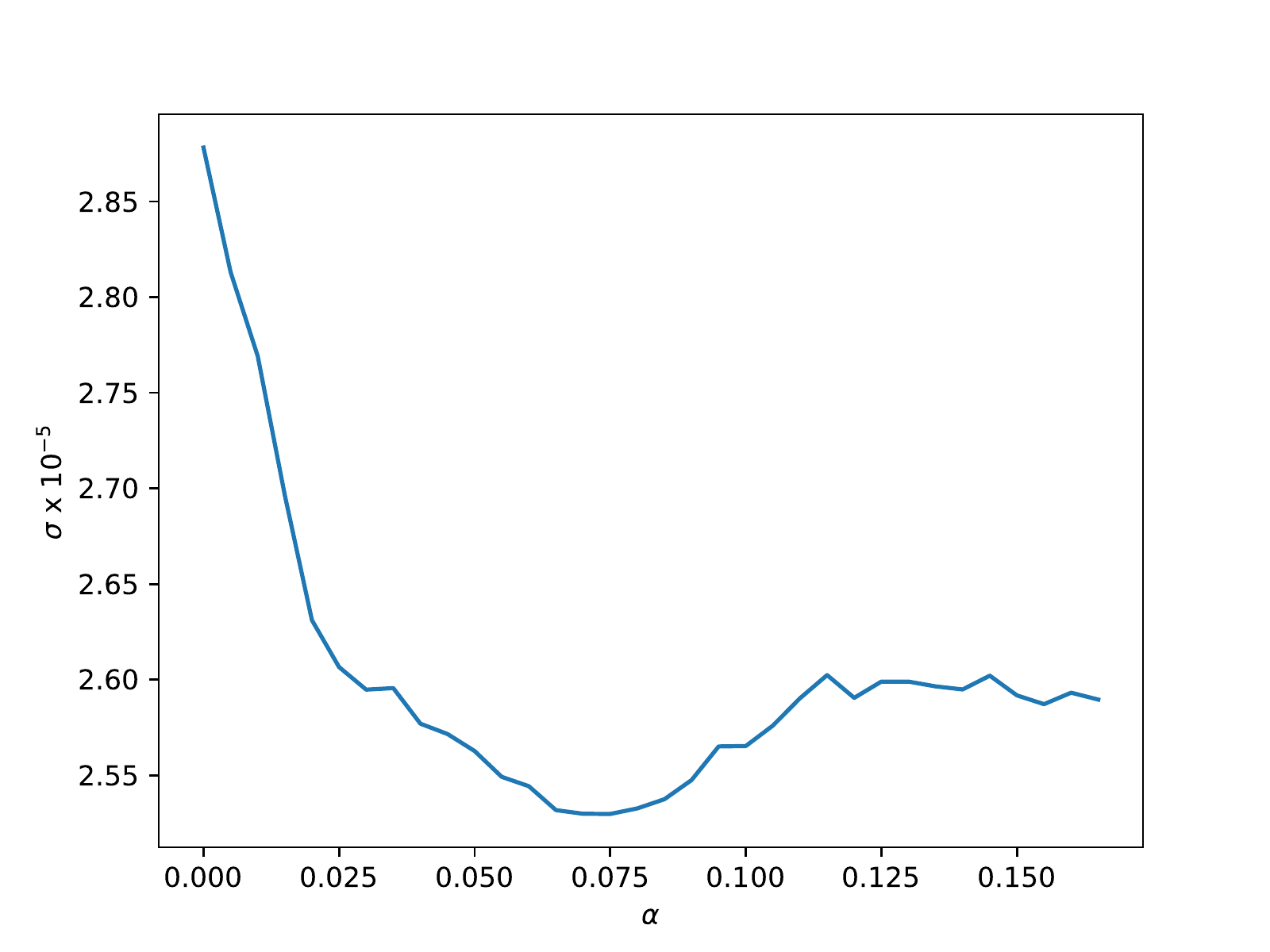}
    \caption{The deviation $\sigma$ as a function of $\alpha$, where for each $\alpha$ the value for $P_{\mathrm{rot}}$ which minimizes $\sigma$ is used.}
    \label{alpha}
\end{figure}

The uncertainty of the differential rotation is difficult to estimate quantitatively. In Fig.~\ref{alpha} we show how $\sigma$ changes as a function of $\alpha$, where the best $P_{\mathrm{rot}}$ value is always chosen for each $\alpha$. The minimum of $\sigma$ is at $\alpha=0.075$, but with a larger $\alpha$, the value of $\sigma$ increases only a little. We did thus not estimate any formal error bars for $\alpha$. This demonstrates further the difficulties in accurately estimating the amount of differential rotation, even for very high-quality data.

From Table \ref{devs} we see that the case with differential rotation is a better fit to the data than the subset 4-6 without differential rotation (note, however, that night 6 has slightly lower S/N than the other nights, which could somewhat increase the value of $\sigma$ for data sets where this night is included), but subsets 1-3 and 2-5 fit the data better than the model with differential rotation. A clear distinction between the cases is thus quite difficult. 

The results above were derived from only Stokes V data. We performed a similar study on I data as well. The results from the I data, however, indicate a very strong anti-solar-like differential rotation (negative $\alpha$). This is probably artificially caused by the cool polar spot, which has a similar influence on the mean line profile as anti-solar differential rotation would have \citep{hackman2019}. Tikhonov regularization prefers the solution with the anti-solar differential rotation, since it tries to minimize the temperature gradient. From Fig.~\ref{profile} one can see that the model for the I profiles is categorically deeper than the observations, which also indicates that the I data is not modeled as well as the V data. Thus, we do not conclude anything based on the differential rotation analysis of the I data, except that the apparent results from I and V data can be radically different. A $\sigma$ map for the I data is shown in Appendix \ref{i-data}.

\begin{figure*}
  \centering
  \includegraphics[bb=30 200 800 340,width=\textwidth,angle=180]{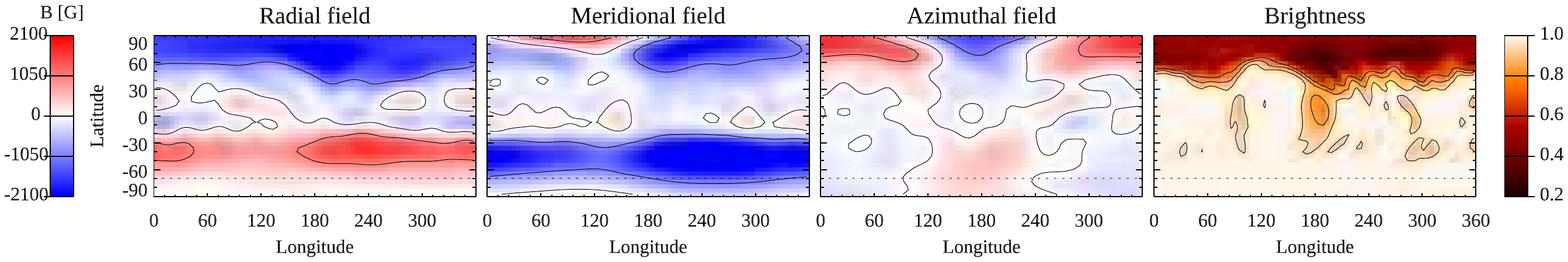}
  \includegraphics[bb=30 200 800 340,width=\textwidth,angle=180]{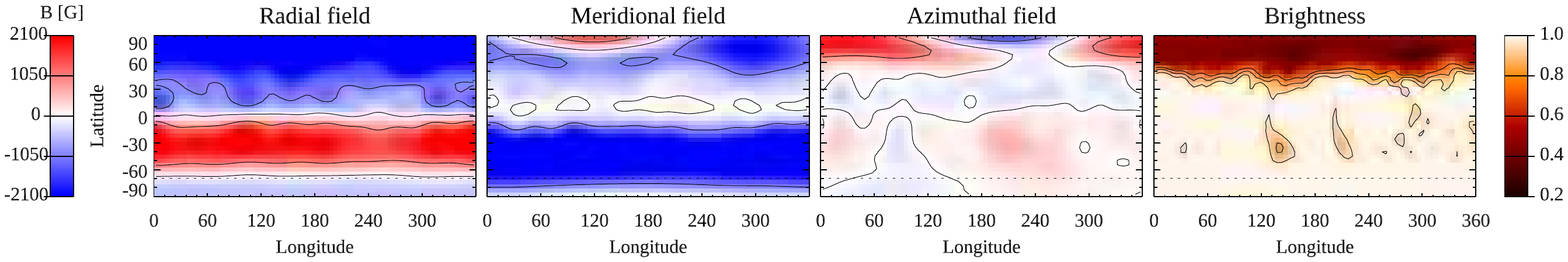} 
   \includegraphics[bb=30 200 800 340,width=\textwidth,angle=180]{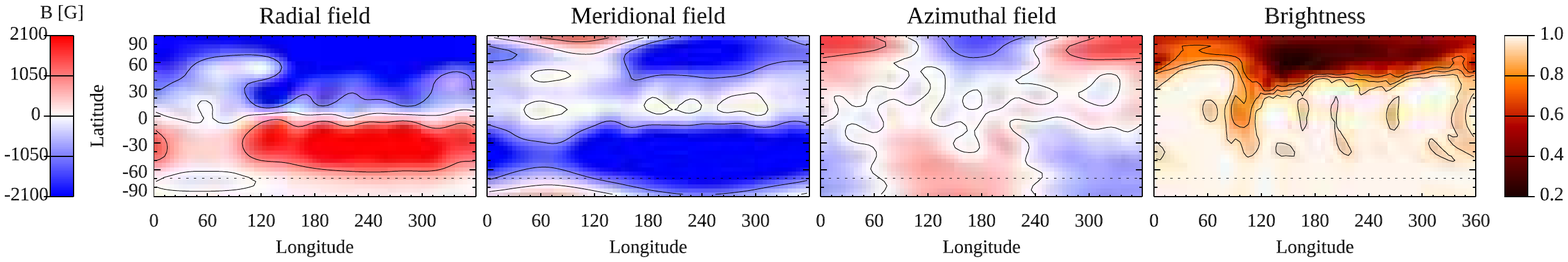}
   \caption{Magnetic field and brightness maps with differential rotation parameters $\alpha = 0.075$ and $P_{\mathrm{rot}}=1.31$ d, for subsets with nights 1-3 (up), nights 2-5 (centre) and nights 4-6 (down). The observed phases are not indicated, since with differential rotation the corresponding longitude is not trivial. The horizontal line indicates the limit of the visible surface due to the inclination. All maps are shown for the epoch HJD=2455699.0}
    \label{maps_difrot2}
\end{figure*}

\begin{table}
\centering
\caption{Deviation between the model and observations $\sigma$ for different data sets and different differential rotation parameters for Stokes V data.}
\label{devs}
\begin{tabular}{c c c c}
\hline\hline
Nights included & $\alpha$ & $P_{\mathrm{rot}}$ & $\sigma \times 10^{-5}$ \\
\hline
1-6 & 0 & 1.3300 & 3.3560 \\
1-3 & 0 & 1.3300 & 1.9456 \\
2-5 & 0 & 1.3300 & 2.1631 \\
4-6 & 0 & 1.3300 & 2.8015 \\
1-6 & 0.075 & 1.31 & 2.5298 \\
1-3 & 0.075 & 1.31 & 1.9440 \\
2-5 & 0.075 & 1.31 & 2.0933 \\
4-6 & 0.075 & 1.31 & 2.6936 \\
\hline
\end{tabular}
\end{table}

\subsection{Case 3: Abruptly evolving magnetic field and differential rotation}

\noindent The two scenarios, abrupt evolution of the magnetic field and differential rotation, do not exclude each other. We thus still explore the possibility of implementing them both.

With the best fit value for the differential rotation, we tried to repeat the analysis keeping the parameters $\alpha=0.075$ and $P_{\mathrm{rot}}=1.31$ d, but now inserting them separately to the subsets 1-3, 2-5 and 4-6 of the data. As is seen from Table \ref{devs}, the inclusion of differential rotation does not much affect $\sigma$ for subset 1-3, while subsets 2-5 and 4-6 now have a clearly better fit. The ZDI maps of the three subsets, including differential rotation, are shown in Fig. \ref{maps_difrot2}. The main features of the maps appear fairly similar as in the maps without differential rotation. Now, though, the magnetic field strength increases from the subset 1-3 to subsets 2-5 and 4-6, while in the maps without differential rotation it decreased from subsets 1-3 and 2-5 to subset 4-6. The corresponding Stokes LSD profiles are shown in Appendix \ref{prof_case3}.


\section{Discussion}

\noindent It is evident that a constant magnetic field without differential rotation cannot explain our data. A comparison of the two different alternatives, a rapidly changing magnetic field or solar-like differential rotation, presented in Table \ref{devs}, reveals that the values for the deviation between model and observations, $\sigma$, do not clearly prefer one alternative over the other. For the subset with nights 1-3, $\sigma$ is very similar with or without differential rotation, whereas for subset 4-6 the solution with differential rotation is preferred. It seems difficult to clearly distinguish between the two cases. The fit to the LSD profiles without magnetic field evolution (Fig.~\ref{prof_difrot}) is fairly good, and the inclusion of differential rotation does also decrease $\sigma$ for the models with magnetic field evolution (although only marginally for nights 1-3). Thus it seems that differential rotation is required to explain the observations. We would argue that the most likely explanation is a situation were differential rotation is present, and some abrupt evolution does indeed also happen in the magnetic field configuration. 
Indeed, if nights 1-3 and 4-6 are modeled independently, including differential rotation, the resulting magnetic field strengths are different, which is the biggest evidence for the abrupt magnetic field evolution to be present besides the differential rotation. At longitude 120, there is
a radial magnetic
field
feature in the 4-6 map and possibly 2-5 map, which is missing in the 1-3 map. If this feature is real, its 
appearance
could also explain the observed difference.

The deviation $\sigma$ increases gradually in all models from nights 1-3 to 2-5 and 4-6. The appearing magnetic region could be an explanation for this, but it could also be explained with the S/N, which is lower for the last nights, especially night 6, and will thus introduce larger residuals between the model and observations. Thus we cannot say with certainty, how significant the possible magnetic field evolution would be.

\section{Implications for ZDI}

\noindent In this section we assume that the hypothesis of the abruptly evolving magnetic field is true, and discuss its possible implications.

Since this data set is of unusually good quality, when compared to other sets collected for V889 Her, it is impossible to say if it is common for the magnetic field of V889 Her to evolve this rapidly. At least this result demonstrates that (Z)DI maps should be regarded with some caution, and the underlying assumption that the magnetic field stays constant through the time span of the observations should be kept in mind. 
For example, the observations in \cite{willamo2019} were generally done over a period of 11 nights, an almost two times longer time span than the six nights analyzed here. Differential rotation was not taken into account in that study either. Its main result, a persistent polar spot, should not be too much affected by abrupt magnetic field evolution or differential rotation, though.

According to the photometry analyzed in \cite{willamo2019}, V889 Her was in a phase of increasing activity in 2011. Around this time, the fairly regular cycle was replaced with a monotonous increase of activity, lasting at least until 2019. It would be reasonable to expect that this kind of abrupt changes in the magnetic field are more common in times of higher activity, such as for the Doppler images published by \cite{willamo2019} during 2012-2017.

The possibility of rapid magnetic field evolution should be kept in mind for other stars as well. Still, it would be reasonable to assume that the magnetic field of V889 Her might change considerably rapidly, since V889 Her is a very active ($\log R'_{\mathrm{HK}}=-4.175$; \cite{lehtinen16}), rapidly rotating star, with a rotation period of approximately 1.3 days. Also in the study by \cite{lehtinen2022_LQHya} of LQ Hya, a similar, almost as rapidly rotating star ($P_{\mathrm{rot}} \approx 1.6$ d), there are a few spectra at close by rotational phases (separated by some stellar rotations), which display some variation in the Stokes V profiles, although much less than in the data analyzed here. In \cite{donati2016,donati2017}, ZDI was applied to the T Tauri star V830 Tau ($P_{\mathrm{rot}} \approx 2.7$ d) using two approximately one month long data sets, separated by one month. In this case, there were slight variations in the magnetic topology, which were assigned to differential rotation, but essentially a stable magnetic surface structure and no apparent problems with ZDI, even for such a long observation span. With less active, more slowly rotating stars, these kind of rapid variations might be less common. 
Nevertheless, this suggests that rapid variations in the magnetic fields of active stars can happen, and more caution should be applied in the interpretation of results from ZDI. Especially if differential rotation is not taken into account, both of these possible effects can affect the results.

\section{Field dynamics}

\begin{table*}
\centering
\caption{The amounts of poloidal and axisymmetric magnetic field, and the average and maximum magnetic field strength for each model. Multiple definitions for the axisymmetry are shown (with modes $m<\ell/2$ and $m=0$ being counted as axisymmetric). The amounts of toroidal and non-axisymmetric magnetic field can be calculated from these as 100\% - the poloidal or axisymmetric component.}
\label{add_params}
\begin{tabular}{c c c c c c c c}
\hline\hline
Nights included & $\alpha$ & $P_{\mathrm{rot}}$ & $E_{\mathrm{pol}}$ [\%] & $E_{m < \ell / 2}$ [\%] & $E_{m=0}$ [\%] & $\langle |B| \rangle$ [G] & $B_{\mathrm{max}}$ [G] \\
\hline
1-6 & 0 & 1.3300 & 94 & 93 & 90 & 1050 & 2720 \\
1-3 & 0 & 1.3300 & 92 & 89 & 86 & 1160 & 3310 \\
2-5 & 0 & 1.3300 & 96 & 98 & 96 & 1450 & 3430 \\
4-6 & 0 & 1.3300 & 80 & 75 & 59 & 540 & 2570 \\
1-6 & 0.075 & 1.31 & 98 & 98 & 97 & 2030 & 4730 \\
1-3 & 0.075 & 1.31 & 92 & 90 & 86 & 1030 & 3350 \\
2-5 & 0.075 & 1.31 & 98 & 98 & 97 & 1910 & 4450 \\
4-6 & 0.075 & 1.31 & 93 & 86 & 83 & 1780 & 4950 \\
\hline
\end{tabular}
\end{table*}

\noindent Regardless of the model for differential rotation or magnetic field evolution, the magnetic field is very much dominated by its poloidal and axisymmetric components. These are listed in Table \ref{add_params}, along with the average and maximum field strength. The magnetic field is notably stronger (average field strength between 1-2 kG in most scenarios\footnote{The magnetic field strength might exceed the limit for the weak field approximation ($\approx$ 1 kG) that is assumed for the LSD profiles \citep[see][]{kochukhov2010_lsd}, which means that the field strength in regions with the strongest field is not reliable. The order of magnitude should, nevertheless, be correct.}) than in the data from 2004 and 2005 from \cite{marsden06} and \cite{jeffers08}, where the field strength was much below 1 kG. Those magnetic field maps also have a much stronger azimuthal component than our maps. Some changes in the magnetic field topology may thus have occurred. During 2004-2005 V889 Her was close to a spot activity maximum, according to the photometric cycle seen in \cite{willamo2019}, but in 2011 the star had entered a state of higher activity than the previous maximum, as its cyclicity had temporarily been replaced with a monotonous decrease in brightness. The state of higher spot activity may thus explain the stronger magnetic field and different magnetic topology.

It should be noted, that the very axisymmetric magnetic field topology makes the estimation of differential rotation more difficult, and could lead to uncertainties regarding it. Differential rotation is also otherwise known to be a difficult parameter to estimate reliably.

The brightness map, again in all the models for differential rotation or magnetic field evolution, is dominated by a high-latitude spot, which is centered a bit off the pole. The low-latitude features are different in the different models, as is the contrast of the high-latitude spot, but the main result from the brightness map, the persistent polar or high-latitude spot, is the same as in earlier Doppler imaging studies of V889 Her \citep[e.g.][]{willamo2019}.

Assuming some correlation between spots and magnetic fields, the possible rapid field evolution would indicate that there are spots appearing and disappearing rapidly on V889 Her. \cite{giles2017} found on their sample of Kepler stars, that hotter stars have shorter spot life times than cooler stars; in their Figure 8, there appear to be spots with short life times, with decay times around 10 days (which they also used as a lower limit for the decay time) only in G and F-type stars, while not in the cooler K and M-stars. As a G-type star, the possible rapid field evolution of V889 Her would fit in this picture.

The most dramatic topological changes, seen in the radial and meridional magnetic field, happen at low latitudes. Note that these changes still take place above the equator -- with the spherical harmonic decomposition of the magnetic field, where the sum of the magnetic field penetrating the stellar surface is forced to be zero, the magnetic features below the equator are quite uncertain. This is because we do not have any information of the magnetic field around the invisible pole, and the inversion forces a strong lower latitude magnetic field with opposite polarity to compensate for the observed magnetic field around the visible pole, even though a large part of the magnetic field is probably unobservable to us.

If the possible magnetic field evolution truly happens close to equatorial regions, as the radial feature around longitude 120 could indicate, then one phenomenon, seen in numerical magnetohydrodynamic simulations, which could possibly give an explanation for this, are \textquoteleft magnetic wreaths\textquoteright. They are 
banded, 
large-scale
magnetic structures, 
which arise 
at low latitudes on both hemispheres
in simulations of rapidly rotating stars \citep{brown2010}. More recently, 
higher-resolution 
simulations 
\cite[see, e.g.,][]{viviani2018}
have revealed that these structures can exhibit small-scale structures, as the toroidal band of magnetic field is twisted and obtains a complicated helical structure. 
This means that it is not only manifested in the azimuthal magnetic field component, but also in the radial and meridional ones. The wreaths show faster and more dynamical behavior than the non-axisymmetric large-scale structures appearing at higher latitudes in these simulations.
Since it is known that there is a large contribution of small-scale fields, which is not captured by ZDI \citep{kochukhov2020}, the rapid time scale of 
the magnetic field evolution supports the conclusion that the actual changes in the magnetic field happen on small scales. Since these wreaths have strong structures of small-scale fields, they could be expected to evolve on short time scales, due to rapid processes such as reconnection and diffusion. Thus, they provide a possible explanation for the fast low latitude changes seen in V889 Her.


\section{Conclusions}

\noindent We have applied ZDI for an extensive set of spectropolarimetry from six consecutive nights in May 2011 for V889 Her. We showed that a constant magnetic field on a solidly rotating star cannot explain the data. Thus we studied two alternative scenarios; rapid evolution of the magnetic field and differential rotation. From our analysis we conclude that it is not easy to distinguish between these two alternatives from ZDI alone. 
It is evident that strong differential rotation is present in V889 Her. The abrupt evolution of the magnetic field is not as certain, but it seems likely that at least some evolution
has
also occurred. Our result  
disagrees with the theory and models where rapidly rotating stars do not show strong differential rotation (compare our $\alpha=0.075$ to e.g. $\alpha < 0.01$ for simulations with comparable rotation \citep{viviani2018}, or $\alpha \lesssim 0.05$ for Kepler stars with $T_{\mathrm{eff}} < 6000$ K in \cite{reinhold2015}). 
We would also urge to caution, when interpreting results of ZDI from long data sets, since it seems possible that some evolution of the large-scale magnetic fields of rapid rotators is possible in short time-scales of only a few days.

\acknowledgments

\noindent Based on observations made with the HARPSpol instrument on the ESO 3.6 m telescopes at the La Silla Observatory under programme ID 087.D-0771(A). T.W. acknowledges financial support of the V\"{a}is\"{a}l\"{a} and von Frenckell foundations. T.H. acknowledges the financial support from the Academy of Finland for the project SOLSTICE (decision No. 324161).
M.J.K-L. acknowledges the support of the Academy of Finland
ReSoLVE Centre of Excellence (grant number 307411).
This project has received funding from the European Research Council (ERC)
under the European Union's Horizon 2020 research and innovation
programme (Project UniSDyn, grant agreement n:o 818665).
O.K. acknowledges support by the Swedish Research Council and the Swedish National Space Agency. We thank the anonymous referee for comments that helped to improve the paper.

\vspace{5mm}

\appendix
\section{Differential rotation for Stokes I data}\label{i-data}
\begin{figure}
  \centering
  \includegraphics[width=9cm]{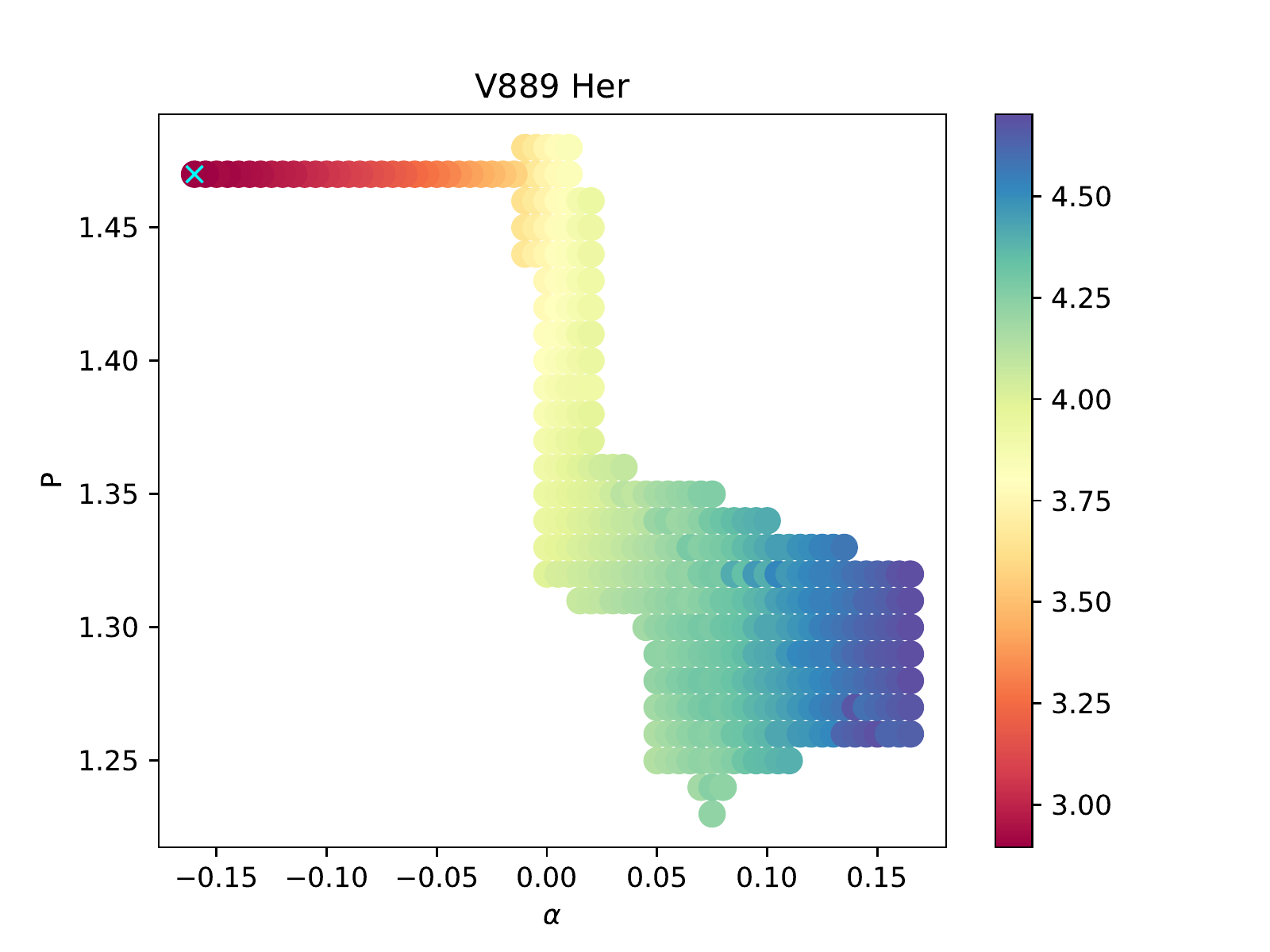}
  \caption{Map of the deviation $\sigma$ for differential rotation parameters of V889 Her, using Stokes I data. The colour scaling corresponds to the deviation between the model and observations in the respective ZDI map, with the unit being $10^{-4}$. The best fit is indicated with a cross.}
  \label{chi2-i}
\end{figure}

\section{Stokes profiles for the model with both differential rotation and magnetic field evolution}\label{prof_case3}
\begin{figure}
    \centering
    \includegraphics[bb=50 50 340 600,width=5cm]{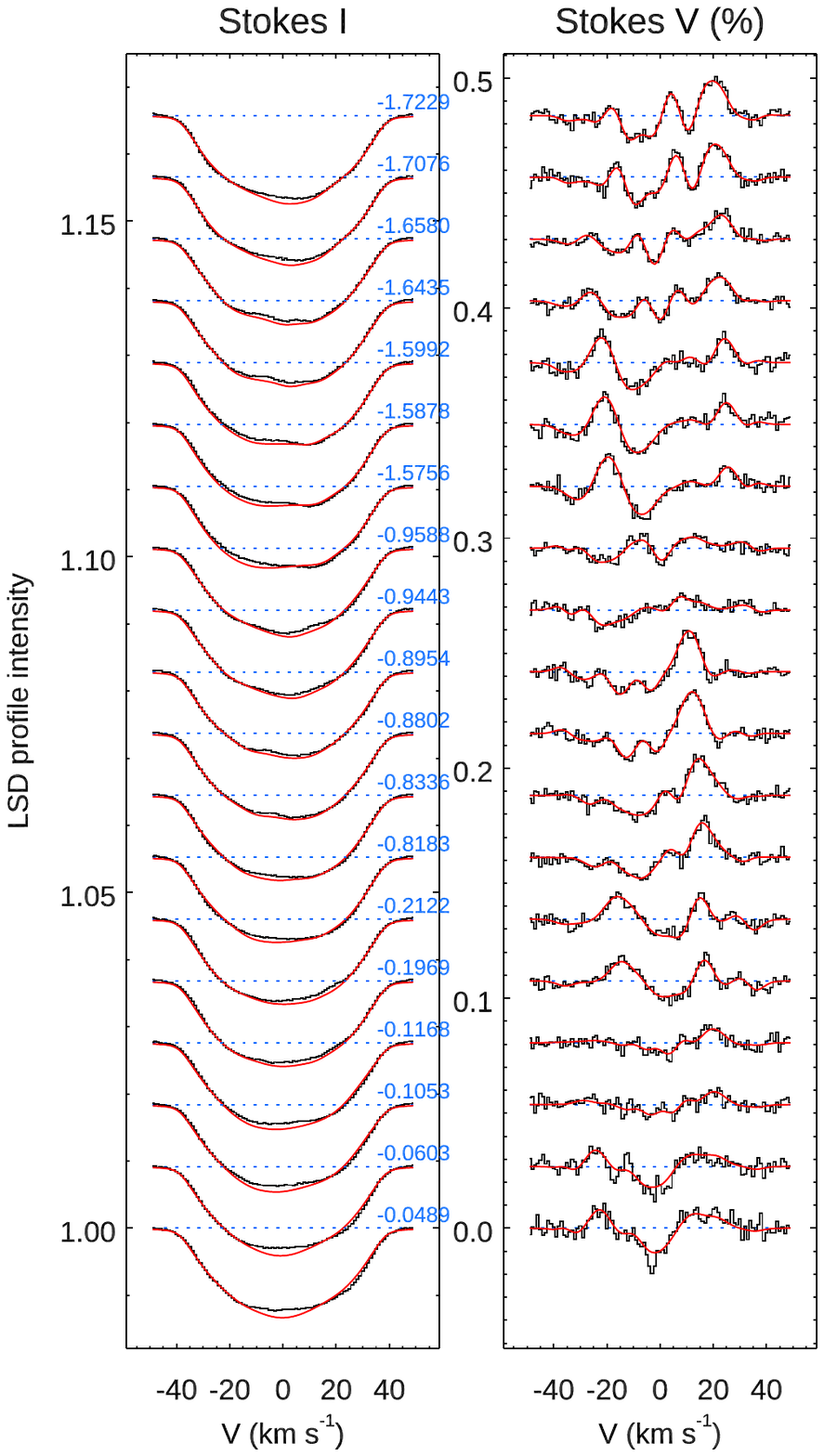}
    \includegraphics[bb=50 50 340 600,width=5cm]{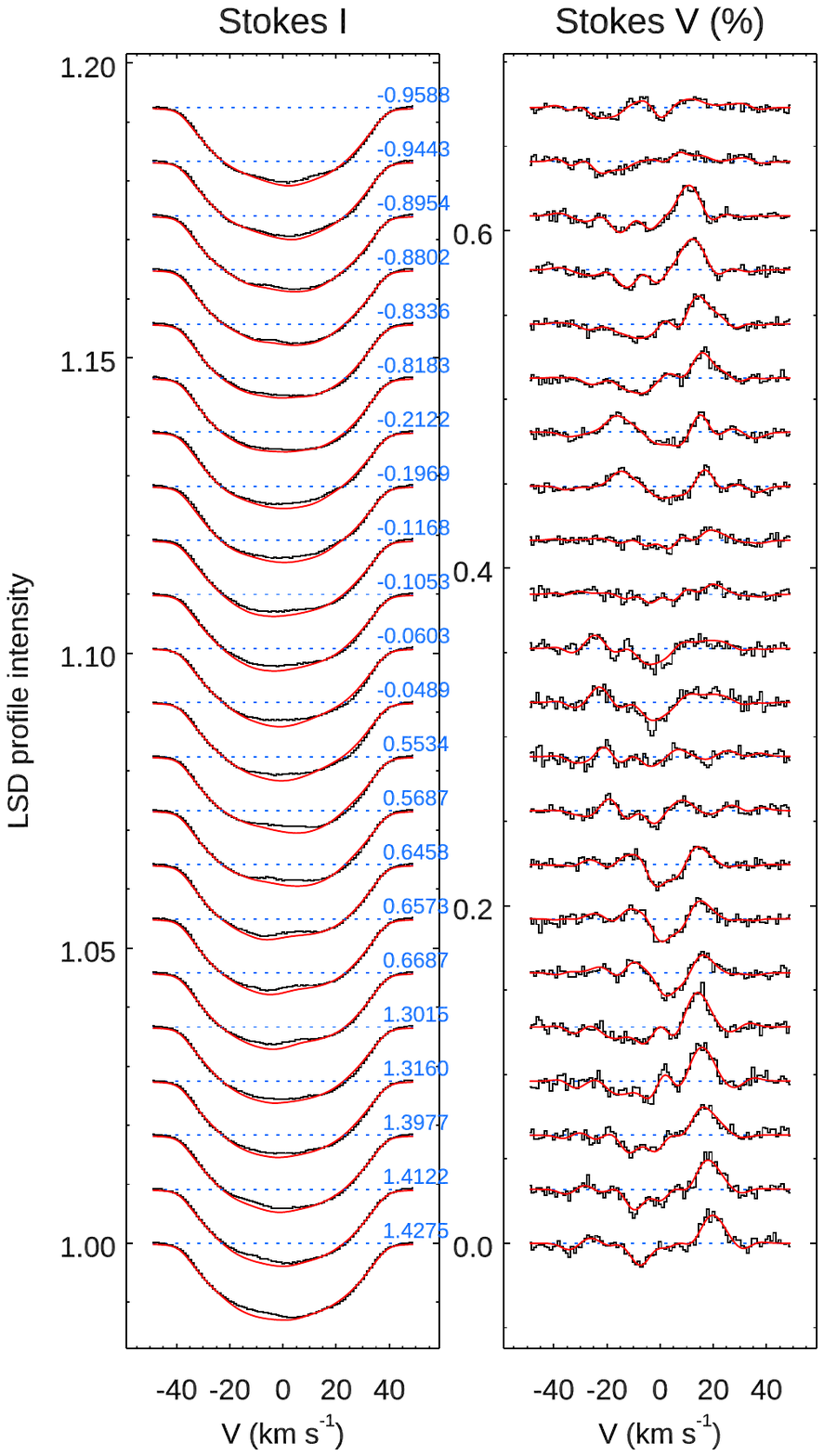}
    \includegraphics[bb=50 50 340 600,width=5cm]{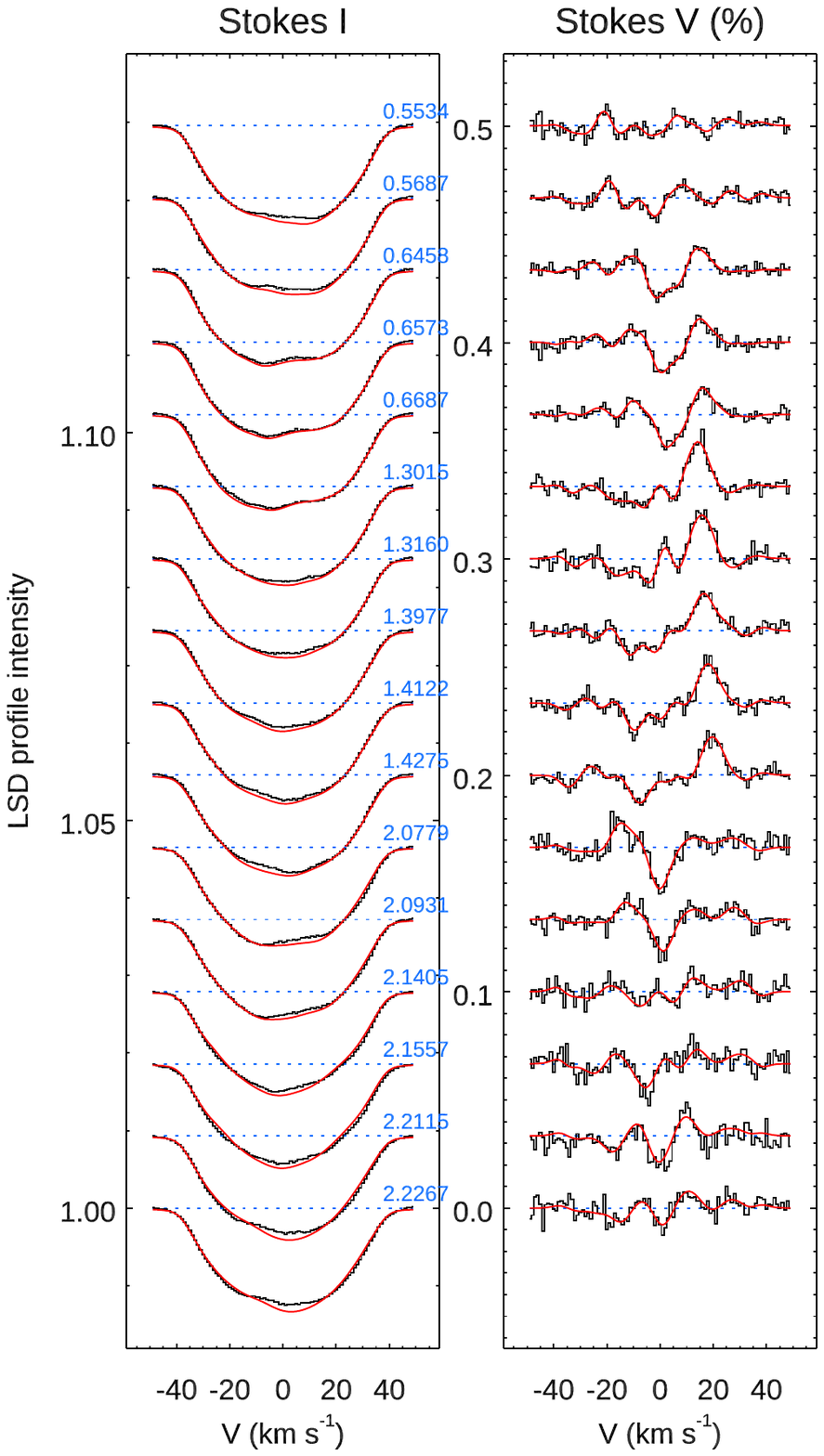}
    \caption{The Stokes profiles for the best value for differential rotation, $\alpha=0.075$ and $P_{\mathrm{rot}}=1.31$ d, when the magnetic field is also allowed to evolve. These are independently modeled for nights 1-3 (left), nights 2-5 (centre) and nights 4-6 (right). The black lines are the observed Stokes LSD profiles, and the red lines are fits made to these.}
    \label{prof_difrot_ev}
\end{figure}


\bibliography{V889Her}{}
\bibliographystyle{plainnat}

\end{document}